\documentclass[reprint,floatfix, aps,prc, twocolumn,notitlepage,
nofootinbib,nobibnotes,
amsmath,amssymb,preprintnumbers]{revtex4-1}
\usepackage{graphicx,xcolor}
\usepackage{physics}
\usepackage{enumitem}   
\usepackage{dcolumn}
\usepackage{hyperref}
\usepackage[normalem]{ulem}
\usepackage{mathtools}
\usepackage{graphicx}
\usepackage{bm}
\usepackage{subfigure} 
\usepackage{xcolor}
\usepackage{float}
\usepackage{placeins}
\usepackage{url}
\usepackage{latexsym}
\usepackage{amsmath}
\usepackage{epsf}
\usepackage{etoolbox}

\makeatletter
\appto\abstract{%
  \let\latexlist\list
  \def\list{\edef\keeprightskip{\the\rightskip}\latexlist}%
  \patchcmd\latexlist{\ignorespaces}{\rightskip\keeprightskip\ignorespaces}{}{}%
}
\makeatother

\begin{document}
\title{Temperature and net baryochemical potential dependence of $\eta/s$ in a hybrid approach }
\author{Niklas G\"otz$^{1,2}$  and Hannah Elfner$^{3,1,2,4}$}
\affiliation{$^1$Goethe University Frankfurt, Department of Physics, Institute for Theoretical Physics, Frank-
furt, Germany}
\affiliation{$^2$Frankfurt Institute for Advanced Studies, Ruth-Moufang-Strasse 1, 60438
Frankfurt am Main, Germany}
\affiliation{$^3$GSI Helmholtzzentrum f\"ur Schwerionenforschung, Planckstr. 1, 64291
Darmstadt, Germany}
\affiliation{$^4$Helmholtz Research Academy Hesse for FAIR (HFHF), GSI Helmholtz Center,
Campus Frankfurt, Max-von-Laue-Straße 12, 60438 Frankfurt am Main, Germany}

\date{\today}

\begin{abstract}
\begin{description}
\item[Background] The effect of a predicted non-constant $\eta/s$($\mu_B$) is largely unexplored in hydrodynamic simulations. Previous studies focus only on a temperature dependence or even only a constant effective shear viscosity.
\item[Purpose] Study qualitative impact of the net baryochemical potential dependence of the shear viscosity to entropy density ratio $\eta/s$ in hydrodynamical simulations.
\item[Method] The effect of a generalized $\eta/s$($T,\mu_B$) is investigated in the hybrid approach SMASH-vHLLE, composed of the hadronic transport approach SMASH and the (3+1)d viscous hydrodynamic code vHLLE. In order to reduce the bias of the result on the equation of state used in the hydrodynamic part of the model, $\eta/s$ is parameterized directly in the energy density and net baryon number density. The parameterization takes into account the constraints of matching to the transport coefficients in the hadronic phase, as well as pQCD results. 
\item[Results] This work demonstrates impact of the density dependence for different system sizes and energies and compares the observables, including yield, mean transverse momentum and integrated elliptic flow, with experimental results in the RHIC - BES region  $\sqrt{s_{NN}}=7.7 - 39.0$ GeV, as the effect of this generalization is especially relevant for intermediate collision energies, for which the system is in equilibrium for a relevant amount of time, but the net baryochemical potential does not vanish.
\item[Conclusions] The effect of an explicit net baryon number dependence on the elliptic flow is negligible and only relevant in the early stages of the collision. Additionally, we find that the proposed parameterization in energy density could be a good proxy for the shear viscosity in the non-equilibrium hadronic transport stage, as the elliptic flow is insensitive to the switching criterion in the range of $\epsilon_{\text{switch}}$=0.1 - 0.5 GeV.
\end{description}
 
\end{abstract}

\maketitle
\section{Introduction}
Collisions of atomic nuclei at relativistic velocities are one of the most important means to study and understand the fundamental properties of matter \cite{Busza_2018}. Different experimental facilities exist to conduct such heavy-ion collisions, such as the Large Hadron Collider (LHC) at CERN, SIS-18 at GSI and the Relativistic Heavy-Ion Collider (RHIC) at BNL, which allow to study the different regions of the QCD diagram. While at high beam energies high temperatures are reached at vanishing net baryon chemical potential, the region accessed by collisions at intermediate energy is of special interest as signatures of of a first order phase transition and a critical end point are expected there \cite{Stephanov_1998,KUMAR_2013}. The theoretical understanding of the dynamics at such intermediate beam energies requires the treatment of finite net baryon chemical potential.

Hybrid approaches based on relativistic viscous fluid dynamics for the hot and dense stage and hadronic transport for the non-equilibrium rescattering have been established as a great tool to describe the dynamics of high-energy heavy-ion reactions. Recent progress has been made in the theoretical description of heavy-ion collisions at finite baryon densities \cite{hybrid,Karpenko_2015,Petersen_2008,Wu_2022,Shen_2017,akamatsu2018dynamically,du20203+, nandi2020constraining} employing such hybrid approaches, which combine the successful description of hadronic transport at low energies, where hadronic interactions prevail, with the high energy description of relativistic hydrodynamic calculations. The hadronic transport typically serves as a hadronic afterburner, which has shown to successfully reproduce experimental observables \cite{Petersen_2014}, but it can also serve to create the initial conditions. Besides the initial conditions, relativistic viscous hydrodynamic calculations need as an input an equation of state as well as the transport coefficients. This indicates also the main advantage of hydrodynamics, namely direct access to the properties of the medium and a potential phase transition.

Regarding the transport coefficients, extensive comparisons of theoretical calculations with experimental flow measurements have shown that there is evidence that the shear and bulk viscosity are not vanishing in the quark-gluon plasma \cite{Shen_2021}. From a theoretical perspective, the transport coefficients are currently not easily accessible from first principle lattice QCD calculations due to numerical challenges \cite{Petreczky_2006,aarts2002transport}. Nevertheless, many theoretical predictions support a non-vanishing shear viscosity over entropy ratio $\eta/s$ \cite{Kovtun:2004de,Auvinen:2017fjw,Nakamura_2005}.

All known fluids in nature have a temperature dependent $\eta/s$, with a minimum close to the phase transition \cite{chapman1990mathematical}. A similiar behavior is also predicted for the shear viscosity of nuclear matter\cite{JETSCAPE:2020shq,JETSCAPE:2020mzn,Ghiglieri:2018dib,Auvinen:2020mpc,Nijs:2020ors,Nijs:2020roc,greiner2011hagedorn,Gorenstein_2008,csernai2006strongly}. Apart from this, there is also theoretical support for a dependence on the net baryochemical potential \cite{denicol2013fluid,itakura2008shear,Kadam_2015,Demir_2009}. Although studies including a non-constant $\eta/s$($T, \mu_B$) in hydrodynamic simulations exist \cite{McLaughlin_2022}, the effect on the observables is largely unexplored. Previous studies focus on a temperature dependence or even a constant effective shear viscosity \cite{Petersen_2008,Petersen_2014,JETSCAPE:2020shq,JETSCAPE:2020mzn,Nijs:2020roc,Auvinen:2017fjw,magdy2021model,2016,Dubla:2018czx}. Therefore, in the following the effects of including both a temperature and a net baryochemical potential dependence of the shear viscosity on the evolution and observables will be investigated.

This work is structured as follows: In Sec. \ref{sec:model}, the hybrid approach {SMASH-vHLLE-hybrid}, within which this study is performed, is briefly summarized. In the following, a parameterization for $\eta/s(T,\mu_B)$ is proposed and the choice of parameters as well as parameterizations chosen for comparison are explained. Additionally, the setup of the computations to investigate the qualitative impact of the $\mu_B$ dependence is illustrated in the same section. In Sec. \ref{sec:results}, this qualitative impact is studied both at the example of the midrapidity yield and mean transverse momentum as well as the elliptic flow. This is accompanied by a comparison to data as well as a study of the effect of changing the switching energy density, which shows that the proposed parameterization approximates the shear viscosity in the hadronic phase. To conclude, a brief summary and outlook can be found in Section \ref{sec:Conclusion}.

\section{Model Description}\label{sec:model}

The theoretical calculations presented in this work are performed using the {SMASH-vHLLE-hybrid} hybrid approach \cite{hybridurl}, which is publicly available and suited for the theoretical description of heavy-ion collisions between $\sqrt{s_{NN}}$ = 4.3 GeV and $\sqrt{s_{NN}}$ = 5.02 TeV. {SMASH-vHLLE-hybrid}  has been shown to reproduce experimental data well across a wide range of collision energies and conserves all charges ($B, Q, S$). It is especially successful in reproducing the longitudinal baryon dynamics at intermediate collision energies \cite{hybrid}.

In the following, a short overview of its components is given (a more detailed description can be found in \cite{hybrid}), as well as the description and motivation of the proposed parameterization for the shear viscosity over entropy density ratio.

\subsection{{Hybrid approach}}
Hybrid approaches combine viscous fluid dynamics for the hot and dense phase of heavy ion collisions with microscopic non-equilibrium transport approaches for the cold and dilute stage. The {SMASH-vHLLE-hybrid} employs the same hadronic transport approach {SMASH} \cite{Weil_2016,dmytro_oliinychenko_2021_5796168,smashurl} for both the initial conditions and the afterburner. This allows for direct comparisons between hydrodynamic and non-equilibrium evolution as well as a gradual turn on of the fluid dynamic stage when moving to higher beam energies. 

\subsubsection{{Hadronic transport}}
{SMASH} effectively solves the relativistic Boltzmann equation by simulating the collision integral through formations, decays and elastic scatterings of hadronic resonances, for which all hadrons listed by the PDG up to a mass of 2.35 GeV are included \cite{ParticleDataGroup:2018ovx}. For hadronic interactions at high energies, hard scatterings are carried out within {Pythia 8} \cite{Sj_strand_2008} and a soft string model is employed. 

The initial conditions of the hydrodynamic evolution are generated by {SMASH} on a hypersurface of constant proper time $\tau_0$, which corresponds to the passing time of the two nuclei \cite{Karpenko_2015}, following the assumption that at this time local equilibrium is reached and the hydrodynamic description becomes applicable \cite{Oliinychenko:2015lva,Inghirami:2022afu}.

In the last stage of the hybrid evolution, {SMASH} is employed for hadron rescattering. The hadrons  obtained from particlization on the Cooper-Frye hypersurface are propagated back to the earliest time and appear subsequently in the hadronic transport evolution and scatter or decay. The calculation is terminated when the medium becomes sufficiently dilute.

\subsubsection{Hydrodynamic evolution}
{vHLLE} \cite{Karpenko_2014} is a 3+1D viscous hydrodynamics code and used to model the evolution of the hot and dense fireball. It solves the hydrodynamic equations
\begin{equation}
    \partial_\nu T^{\mu \nu} =0 \quad, 
\end{equation}
\begin{equation}
    \partial_\nu j^\nu_B=0 \quad \partial_\nu j_Q^\nu=0 \quad \partial_\nu j_S^\nu=0 \quad .
\end{equation}
These equations represent the conservation of net-baryon, net-charge and net-strangeness number currents as well as the conservation of energy and momentum. The energy-momentum tensor is decomposed as 
\begin{equation}
    T^{\mu\nu}=\epsilon u^\mu u^\nu -\Delta^{\mu \nu}(p+\Pi)+\pi^{\mu\nu}
\end{equation}
with $\epsilon$ the local rest frame energy density, $p$ and $\Pi$ the equilibrium and bulk pressure and $\pi^{\mu\nu}$ is the shear stress tensor. These equations are solved in the second order Israel-Stewart framework \cite{Denicol_2014,Ryu_2015}.

In order to initialise the hydrodynamic evolution with the iso-$\tau$ particle list generated from {SMASH}, as outlined in the last section, some smoothing has to be performed in order to prevent shock waves. For this purpose, a Gaussian smearing kernel \cite{Karpenko_2015} with the parameters taken from \cite{hybrid} is applied. At this step, particles have been converted into fluid elements which are evolved using a chiral model equation of state \cite{Steinheimer_2011} matched to a hadron gas equation of state. This evolution is performed until a switching energy density is reached, which is set in this work to a default value of $\epsilon_{\text{switch}}$=0.3 GeV/fm$^3$,  if not mentioned differently.  At this point, the freezeout hypersurface is constructed using the {CORNELIUS} subroutine \cite{Huovinen_2012} and the thermodynamical properties of the freezeout elements are calculated according to the {SMASH}  hadron resonance gas equation of state \cite{hybrid} to prevent discontinuities. $\epsilon_{\text{switch}}$ is a free parameter and different choices are possible. This relies on the assumption that there is a region in the QCD phase diagram where both hydrodynamic evolution as well as hadronic transport are equivalently applicable, which in turn validates the application of hybrid approaches. It is important to notice that $\epsilon_{\text{switch}}$ is a technical parameter and does only control the application of the hybrid and transport approach, but does not determine which degrees of freedom are realized in the QCD matter.

\subsubsection{{Particle sampler}}

As {SMASH} evolves particles and not fluid elements, particlization has to be performed, for which the {SMASH-vhLLE} hybrid approach applies the {SMASH-hadron-sampler} \cite{samplerurl}. The {SMASH-hadron-sampler} employs the grand-canonical ensemble in order to particlize each surface element independently. Hadrons are sampled according to a Poisson distribution with the mean at the thermal multiplicity, and momenta are sampled according to the Cooper-Frye formula \cite{cooper1975landau}. The corrections to the distribution function $\delta f_{\rm shear}$ associated to a finite shear viscosity are considered using the Grad's 14-moment ansatz, for which we assume that the correction is the same for all hadron species. Due to ($T,\mu_B$) dependence of the shear viscosity, the corrections also have this dependence implicitly. This procedure provides a particle list that can be evolved by {SMASH}. Because of the grand-canonical sampling procedure, quantum numbers are not conserved event-by-event but only on average. This is sufficient for this work as we are only interested in averaged observables. It is crucial to match the hadronic degrees of freedom in the equation of state to the ones that are active in SMASH to maintain energy conservation at the transition even for the average over events. A more thorough introduction into the sampling procedure can be found in \cite{Karpenko_2015}.

\subsubsection{Configuration details}
For each simulation presented, 100 event-by-event viscous hydrodynamics events were generated, each initialized from one single {SMASH} initial condition. The smearing parameters for the transition from transport to hydrodynamic evolution are taken from Table 1 of \cite{hybrid}, as these have been shown to reproduce experimental data well. From the resulting freezeout hypersurface 1000 events are sampled for the hadronic afterburner evolution, in order to allow for on-average quantum number conservation. Whereas the bulk viscosity is set to zero for all runs, the choice of the shear viscosity will be discussed in the following.
Unless stated otherwise, the code versions {SMASH-vHLLE-hybrid:03232b2}, {SMASH-2.1.4}, {vhlle-params:99ef7b4} and {SMASH-hadron-sampler-1.0} are used in this work. The modifications to the hydrodynamic code in order to allow for different parameterisations are based on {vHLLE:efa9e28}.

\subsection{Parametrization of $\eta/s$}
\begin{figure}[t]
    \centering
    \includegraphics{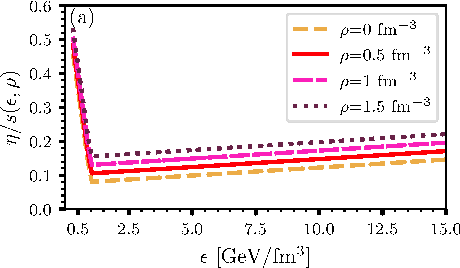}
    \includegraphics{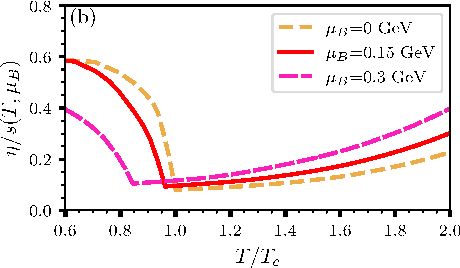}
    \caption{$\eta/s(\epsilon,\rho)$ for $S_\rho=0.05$ fm$^3$ (top) and the same parametrization mapped into $(T, \mu_B)$ using a chiral equation of state (bottom).}
    \label{fig:para}
\end{figure}

This work aims at investigating both a temperature and net baryochemical potential dependence of the shear viscosity in the fireball. It is a common approach to parametrize $\eta/s$ directly in $T$ and it would be possible to add additional terms proportional to $\mu_B$ as well. However, parametrizing instead in the local rest frame energy density $\epsilon$ and the net baryon number $\rho$ has the advantage that these are the quantities which are evolved throughout the hydrodynamic evolution. Due to this, the parameterization is independent of the equation of state, even if the shear viscosity itself still depends on the equation of state due to the entropy density in the denominator. Additionally, as outlined before, a minimum is expected close to the cross-over transition between the quark-gluon plasma and the hadron gas phase. As the transition line is expected to follow approximately a line of constant energy density, this behavior is easier to replicate with a parameterization in the energy density itself.

For this qualitative study, several simplifications are made. First of all, although at finite $\mu_B$ not $\eta/s$ but $\eta T/w$, with the enthalpy $w=\epsilon +p$, is the more correct measure of fluidity in the system \cite{koch_2010}, we use $\eta/s$ as an approximation, as both terms have the same limit for small $\mu_B$. \
Next, as pointed out above, we expect a minimum of $\eta/s$ near the transition. In the limit of vanishing $\mu_B$, we want to replicate this behavior of $\eta/s (T)$ also in $\eta/s (\epsilon)$, as $\epsilon \sim T^4$. As observables are more sensitive to the effective shear viscosity than to the functional form of the shear viscosity \cite{Gardim_2021}, we restrict ourselves to a linear dependence of the shear viscosity both in the region of high and low energy densities. Up to now, this already implies a $\mu_B$-dependence, since our parameterization $\eta/s(\epsilon)$ is dependent on $\mu_B$, whereas $\eta/s(T)$ is not. For example, the minimum of $\eta/s(\epsilon)$ moves to lower temperatures with increasing $\mu_B$. Nevertheless, we want to be able to study the effects of a $\mu_B$-dependence more in detail, which is why we include a linear term proportion to the net baryon number density $\rho$. We chose the net baryon density over the baryon density as this is the quantity which is evolved in the hydrodynamic evolution. Although negative values for the net baryon density are allowed, in practise, due to smoothing out the initial condition and the significant values of $\mu_B$ at the collision energies we have studied, the net baryon density is greater equal zero for all cells of the hydrodynamic evolution. Together with the condition that the shear viscosity always stays positive, this leads us to the following functional form:
\begin{widetext}
\begin{equation}
    \eta/s (\epsilon, \rho) = \max\left(0,  (\eta/s)_{\text{kink}}  + \begin{cases}
S_{\epsilon, H}(\epsilon - \epsilon_{\text{kink}}) + S_{\rho}\rho, & \epsilon < \epsilon_{\text{kink}}\\
S_{\epsilon, Q}(\epsilon - \epsilon_{\text{kink}}) +S_{\rho}\rho & \epsilon > \epsilon_{\text{kink}}
\end{cases} \right)
\end{equation}
\end{widetext}
This form of the parameterization results in five parameters, out of which all but $S_\rho$ will be fixed by constraints. We set $\epsilon_{\text{kink}}$, the position of the minimum and vanishing net baryochemical potential, to 1 GeV/fm$^3$ according to lattice QCD \cite{borsanyi2014full} and experimental results \cite{Cleymans_2006}. The value of the shear viscosity at this minimum is set to the KSS-bound \cite{Kovtun:2004de}. The steepness for the high energy density region is motivated by matching pQCD results \cite{Ghiglieri:2018dib} at a temperature of 400 MeV and vanishing net baryochemical potential. The steepness in the low energy region is set to match the shear viscosity extracted from box calculations in {SMASH} at vanishing net baryochemical potential and at the particlization temperature in order to reduce discontinuities. It is however important to note that this is only an approximation of the shear viscosity in hadronic transport, as box calculations differ substantially from the expanding medium in the rescattering phase. It is important to note that in general, this parameterization allows the shear viscosity to take any value greater zero, and also to violate the KSS bound depending on the choice of parameters as this is in general not a strict bound for non-conformal theories\cite{Critelli_2014}.

This leaves one free parameter, $S_\rho$, which is going to be varied in order to investigate its effect on the observables. The parameterization for the choice of $S_{\rho}=0.05$ fm$^3$ can be see in in Fig. \ref{fig:para}, both in ($\epsilon$,$\rho$) as well as ($T$,$\mu_B$), for which the mapping is performed by using the same equation of state as employed in {vHLLE}. Both show the typical kink structure. The temperature dependence of the minimum in ($T$,$\mu_B$) is present also for vanishing $S_\rho$, whereas the value of the shear viscosity over entropy density at the minimum changes only due to non-vanishing $S_\rho$.

In the following, we want to compare our parameterization with other existing choices for constant or temperature-dependent $\eta/s$. The first comparison choice is a constant value for $\eta/s$, with different values at different collision energies (see table 1 in \cite{hybrid}). These are the default values in {SMASH-vHLLE-hybrid}, and were originally taken from {UrQMD+vHLLE} in order to be in optimal agreement with experimental data. Such a choice of $\eta/s$ is often very successful, as many observables are mostly sensitive to the effective shear viscosity \cite{Gardim_2021}, but rather insensitive to different values of the viscosities in different regions and stages of the evolution of the fireball. 

The second parameterization chosen for comparison is representative for the increasing efforts of extracting $\eta/s(T)$ from experimental data by means of Bayesian analysis. We use the parameterization in \cite{JETSCAPE:2020mzn} and choose parameters which lie central in the 60\% confidence interval of the posterior. Though being tuned to be in good agreement with experimental data at high collision energies and providing a minimum of the shear viscosity close to the transition temperature, it does not include any dependence on the net baryochemical potential.

\FloatBarrier
\section{Results}\label{sec:results}

\begin{figure*}
    \centering
    \includegraphics{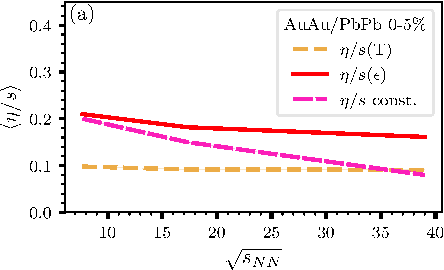}
    \includegraphics{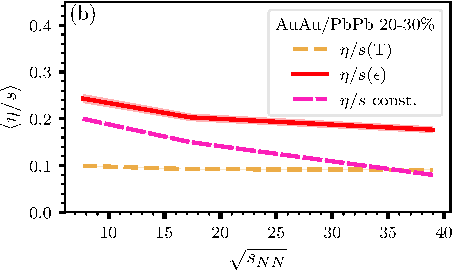}
    \includegraphics{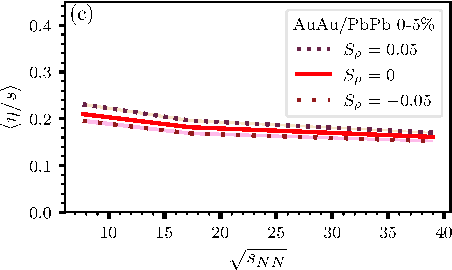}
    \includegraphics{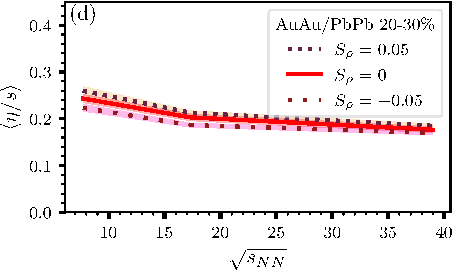}
    \caption{Energy-density weighted mean shear viscosity over entropy ratio throughout the whole hydrodynamic evolution. Top: Comparison between different parametrization choices for central and off-central collisions. Bottom: impact of net baryon number density dependence for central and off-central collisions.}
    \label{fig:meaneta}
\end{figure*}

In the following, we investigate the qualitative effects that the proposed parameterization and different choices of $S_\rho$ have on the evolution and observables in heavy-ion collisions. Since the region of non-vanishing net baryochemical potential is of special interest, the chosen energies and systems are AuAu collisions for $\sqrt{s_{NN}}$= 7.7 and 39 GeV as well as PbPb collisions at $\sqrt{s_{NN}}$= 17.3 GeV.  For observables, the focus will lie on midrapidity yield and mean transverse momentum of central collisions, which are only weakly sensitive to the shear viscosity but can be computed with precision even at low statistics, and the integrated elliptic flow $v_2$ at centrality ranges 20-30\%, which is highly sensitive to the shear viscosity. Centrality is for this study defined by the impact parameter. For the collision setups investigated in this study, the temperature reaches throughout the hydrodynamic evolution values between 108 and 407 MeV, whereas the net baryochemical potential reaches values between 0 and 583 MeV.

These observables will be compared for two sets of parameterizations: One which compares $\eta/s (\epsilon)$ (with $S_\rho$ equal to zero) to both the constant and temperature dependent parameterization mentioned in the last section for comparison to prior choices and another set where the impact of an explicit net baryon number dependence by comparing the cases $S_\rho=$ 0, 0.05 and -0.05 fm$^3$ is studied.

\subsection{Effective $\eta/s$ and time evolution}

An important mean to compare different, non-constant parameterizations of the shear viscosity is the effective shear viscosity, which is a weighted average throughout the hydrodynamic evolution. Different ways to weight the contribution of the shear viscosity in a single fluid cell at a specific time step exist, but for this work, we restrict ourselves to choose the energy density as a weight. The result of this comparison can be seen in Fig. \ref{fig:meaneta}. Here and in all following figures, the standard statistical error is plotted as a band.

We see that the parameterization in the energy density reaches in general higher values for the shear viscosity than the parameterizations chosen for comparison, especially in peripheral collisions. The origin of this lies in the matching to the shear viscosity in {SMASH} which is larger than the one in the purely temperature dependent parameterization.  This hints at the fact that many cells close to the switching energy density provide a significant contribution to the averaged shear viscosity. The temperature dependent shear viscosity parameterization shows only a small dependence on the collision energy, as the posterior of the Bayesian inference is centered around relatively small steepnesses in comparison to the proposed parameterization, while also having a similar value of the minimum. The constant values for the shear viscosity, on the other hand, agree with the effective values for low collision energies but have a much faster decline.
\begin{figure}
    \centering
    \includegraphics{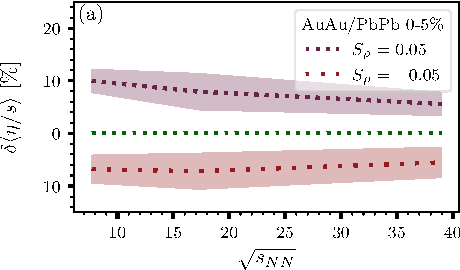}
    \includegraphics{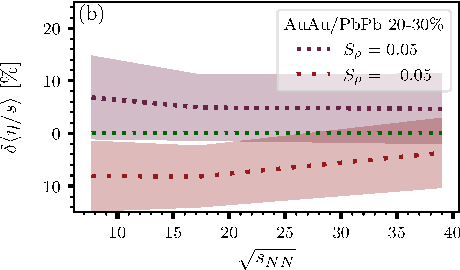}
    \caption{The $S_\rho$-dependent plots of Fig. \ref{fig:meaneta} normalized to the parameterization with $S_\rho=0$, in order to show the deviation in percentage. }
    \label{fig:meaneta_percent}
\end{figure}

For the comparison of $\rho$ dependence, Fig. \ref{fig:meaneta_percent} shows that the effect is rather small and reduces with the collision energies, and, to a smaller degree, with decreasing centrality of the collision. In both cases either the density of the medium is reduced, as either the net baryochemical potential is lower, or the collision is only happening in the more peripheral zones of the nucleus.

\begin{figure}
    \centering
    \includegraphics{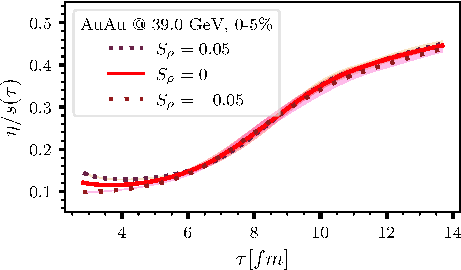}
    \caption{Time evolution of the energy density weighted shear viscosity over entropy ratio at $\sqrt{s_{NN}}=39$ GeV central collisions. The net baryon number density dependence is only relevant for the early stage of the collision.}
    \label{fig:etatime}
\end{figure}

 A closer insight is given in Fig. \ref{fig:etatime}, which shows the effective shear viscosity over entropy ratio during each timestep for central collisions at $\sqrt{s_{NN}}$= 39 GeV. The strongest effect can be seen in the very early stages of the hydrodynamic evolution, where densities are high. In general, the parameterization in $\epsilon$ shows a strong time dependent behavior. At the observed energies, many fluid elements are initialised close or slightly above $\epsilon_{\text{kink}}$ and have small shear viscosities, which quickly decreases when the cells cool down and become more dilute. Let us emphasize here again, that this does not imply that the finite net baryochemical potential is irrelevant, since our $\eta/s(\epsilon)$ incorporates this dependence implicitly. 

\FloatBarrier
\subsection{Bulk observables}

\begin{figure}
    \centering
    \includegraphics{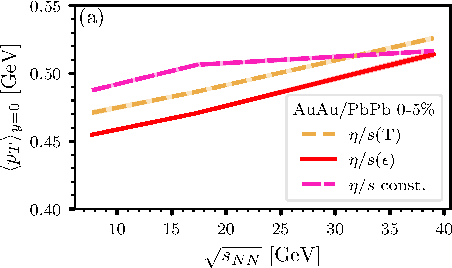}
    \includegraphics{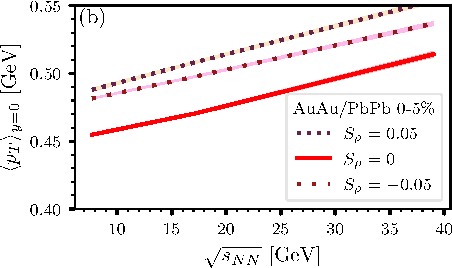}
    \caption{Excitation function of mean transverse momentum of charged hadrons at midrapidity ($|y|<0.5$) for different parametrization strategies (top) and different values of the net baryon number density dependence (bottom).}
    \label{fig:pT}
\end{figure}
A first insight into the effect of net baryochemical potential dependence can be gained by looking at the bulk observables in central collisions. In Fig. \ref{fig:pT}, the mean transverse momentum at midrapidity for charged hadrons is compared for the different parameterizations. The parameterization in $\epsilon$ results in smaller mean transverse momentum for low collisions energies, but approaches the value of other parameterizations at higher values. Introducing a $\rho$ dependence, both positive and negative, increases the transverse momentum, which hints to non-linear effects to the evolution of the fireball.

\begin{figure}
    \centering
    \includegraphics{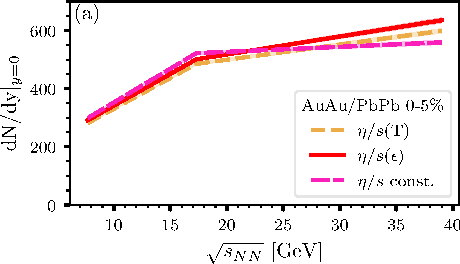}
    \includegraphics{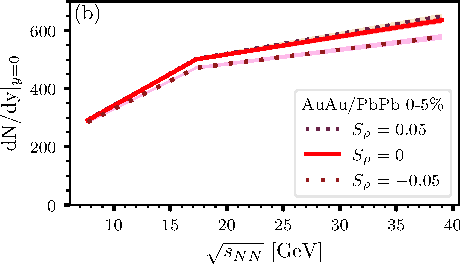}
    \caption{Excitation function of midrapidity yield ($|y|<0.5$) of charged hadrons for different parametrization strategies (top) and different values of the net baryon number density dependence (bottom).}
    \label{fig:y}
\end{figure}

In Fig. \ref{fig:y}, the midrapditiy yield for charged hadrons is shown as a function of beam energy. The differences between the different parametrization strategies become only significant at higher collision energies, where the energy density dependent parameterization increases the yield. With respect to the explicit net baryon number dependence, the addition of $S_\rho$ is also more important at higher collision energies. The effect is however not linear - adding a term which increases the shear viscosity with increasing net baryon number density has virtually no effect on the yield, whereas decreasing the shear viscosity with increasing net baryon number density significantly reduces the yield.

The increasing importance of $S_\rho$ with increasing collision energy seems surprising, as $\mu_B$ decreases with collision energy. However, as observed in \cite{hybrid}, the lifetime of the fireball evolved within hydrodynamics strongly increases in this energy range, resulting in an increasing importance of the choice of parameterization.

\begin{figure}
    \centering
    \includegraphics{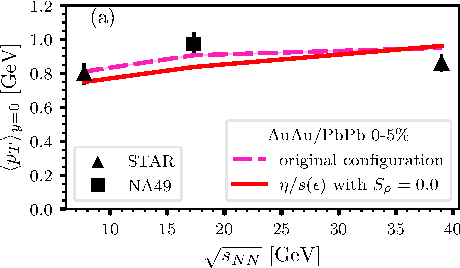}
    \includegraphics{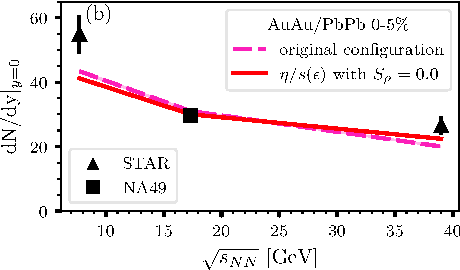}
    \caption{Comparison of proton mean transverse momentum and midrapidity yield to data from NA49 \cite{na49pt} and STAR \cite{Adamczyk_2017}.}
    \label{fig:data_ypt}
\end{figure}

At this point, it is also interesting to take a look at the available experimental data, even though we are aiming only at a qualitative study of the impact of an net baryon chemical potential dependence in the transport coefficients. In Fig. \ref{fig:data_ypt}, the mean transverse momentum and midrapidity yield for protons are compared between the parameterization in the energy density with vanishing $S_\rho$, the original, constant shear viscosity and experimental values. The choice of protons for this comparison is motivated by the available data. In both cases, the agreement is fairly similar and experimental values are reproduced in a reasonable fashion. 

\FloatBarrier
\subsection{Elliptic flow}
\begin{figure}
    \centering
    \includegraphics{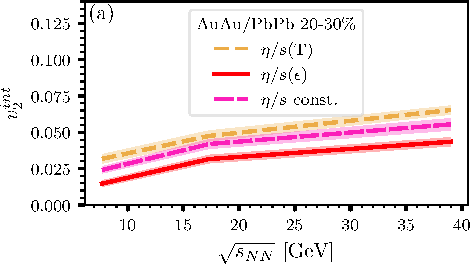}
    \includegraphics{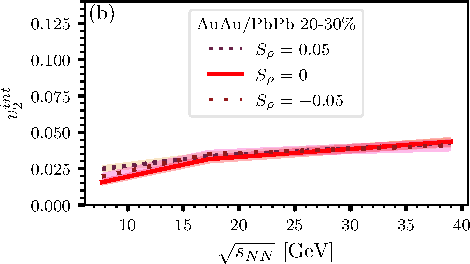}
    \caption{Integrated event plane elliptic flow of charged hadrons at midrapidity ($|y|<0.5$) for different parametrization strategies (left) and different values of the net baryon number density dependence (right).}
    \label{fig:v2}
\end{figure}
The anisotropic flow coefficients are very sensitive to the shear viscosity \cite{magdy2021model}, especially higher order flow coefficients. These are however only accessible with high statistics, which is the reason why we restrict ourselves to the integrated $v_2$ in collisions at 20\%-30\% only. 
The results can be found in Fig. \ref{fig:v2} for the excitation function of $\langle v_2 \rangle$. We notice first of all that the energy density dependent parameterization leads to a significantly reduced elliptic flow in comparison to the investigated alternatives, which can be explained by the higher value of the effective shear viscosity. In contrast the different values of $S_\rho$ have only minimal effects on the flow which are mostly restricted to low collision energies. This might be a hint that the late phases of the hydrodynamic evolution are more important for the flow than the early stages.

\begin{figure}
    \centering
    \includegraphics{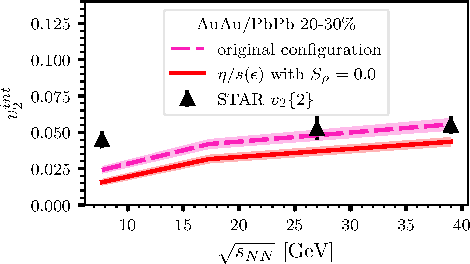}
    \caption{Comparison of charged hadron integrated event plane elliptic flow at midrapidity to STAR data \cite{Adamczyk_2018}.}
    \label{fig:v2_data}
\end{figure}

Comparing the integrated flow of charged hadrons with experimental data, the reduction in flow introduced with $\eta/s(\epsilon)$ increases disagreement with experimental data. Except for low energies, where also the original configuration of {SMASH-vhlle-hybrid} is missing the data point due to the short hydrodynamic evolution, the agreement in the original configuration is better than with $\eta/s(\epsilon)$. This again does not mean that the net baryon chemical potential dependence can be neglected, but is rather a result of our qualitative approach, where we did not attempt to fit other parameters of the approach at the same time. 

\subsubsection*{Impact of switching energy density}
\begin{figure}
    \centering
    \includegraphics{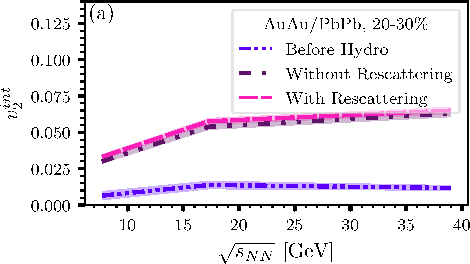}
    \includegraphics{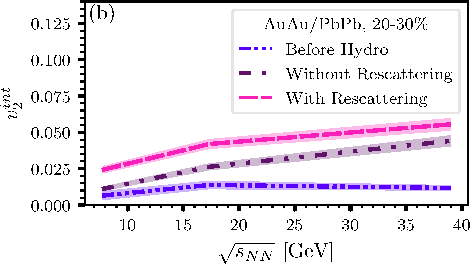}
    \includegraphics{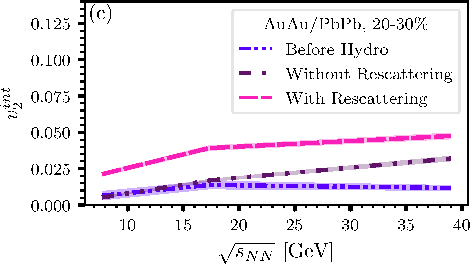}
    \caption{Contribution to the elliptic flow from the different stages of the hybrid simulation, for different values of the $\epsilon_{\text{switch}}$. Top: 0.1 GeV/fm$^3$, middle: 0.3 GeV/fm$^3$, bottom 0.5 GeV/fm$^3$.}
    \label{fig:epsswitch}
\end{figure}
As pointed out before, the duration of the hydrodynamic evolution has an important effect on the role the parameterization of $\eta/s$ for the observables, as the viscous corrections are applied to the system for a longer period. The time spent during the hydrodynamic evolution is defined by the point, when the last fluid element reaches an energy density smaller than $\epsilon_{\text{switch}}$. Depending on the value of $\epsilon_{\text{switch}}$, the system spends more time in the hydrodynamic evolution and less in the transport calculation, or vice versa. This, in turn, also determines in which part of the hybrid approach elliptic flow is generated. Fig. \ref{fig:epsswitch} builds on an investigation performed in \cite{PhysRevC.88.064908} and compares the elliptic flow in the initial condition with the elliptic flow without rescattering, which approximates the flow at the end of the hydrodynamic evolution, and the flow of the full evolution. The calculations were done choosing the default constant shear viscosity and $\epsilon_{\text{switch}}$=0.5, 0.3 and 0.1 GeV/fm$^3$.

The comparison of the flow contribution of the different phases of the hybrid approach shows that for increasing $\epsilon_{\text{switch}}$ and decreasing collision energy, more and more of the flow is generated in the hadronic transport evolution. For $\epsilon_{\text{switch}}$=0.5 GeV/fm$^3$, the contribution of the hydrodynamic evolution is only of significant importance from $\sqrt{s_{NN}}$=39 GeV on, whereas for a value of 0.1 GeV/fm$^3$, the afterburner stage does not contribute anymore to the observed elliptic flow.
\begin{figure*}
\centering
    \includegraphics{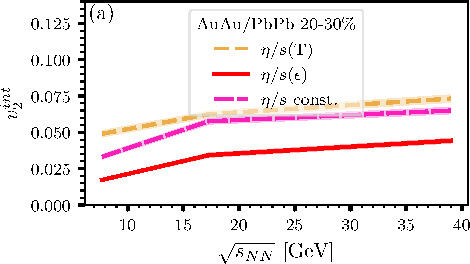}
    \includegraphics{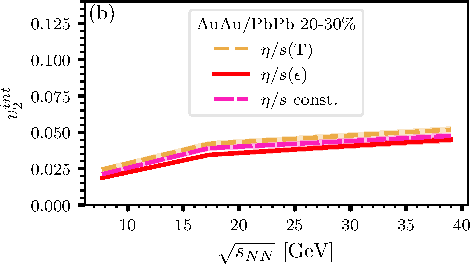}
    \includegraphics{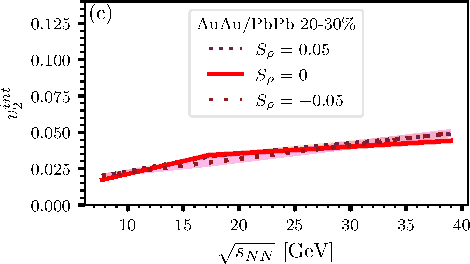}
    \includegraphics{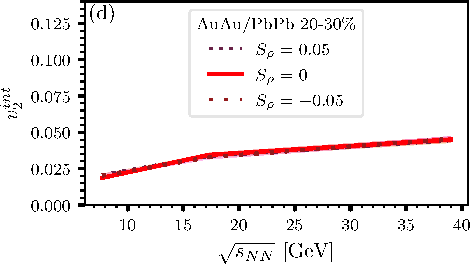}
\caption{Integrated event plane elliptic flow of charged hadrons at midrapidity ($|y|<0.5$) for different parametrization strategies (top) and different values of the net baryon number density dependence (bottom), for $\epsilon_{\text{switch}}$ decreased to 0.1 GeV/fm$^3$ (left) and increased to 0.5 GeV/fm$^3$ (right).}
\label{fig:flowconst}
\end{figure*}
\begin{figure}
\centering
    \includegraphics{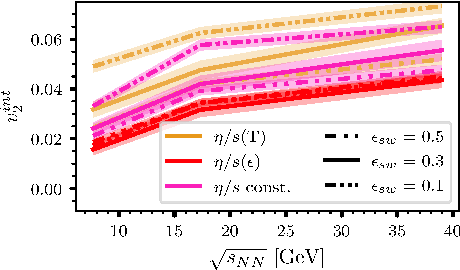}
    \caption{Integrated event plane elliptic flow of charged hadrons at midrapidity ($|y|<0.5$) for different parametrization strategies and values of $\epsilon_{\text{switch}}$.}
    \label{fig:main}
\end{figure}
In light of this insight, Fig. \ref{fig:flowconst} shows the integrated charged hadron elliptic flow for the cases $\epsilon_{\text{switch}}$=0.5 GeV/fm$^3$ and 0.1 GeV/fm$^3$ and should be compared with Fig. \ref{fig:v2}. Independent of $\epsilon_{\text{switch}}$, the impact of $S_\rho$ is virtually independent of the switching energy density. The elliptic flow for $\eta/s(T)$ and constant $\eta/s$ changes significantly when a different value of $\epsilon_{\text{switch}}$ is chosen. On the other hand, this is not the case for $\eta/s(\epsilon)$. 

This becomes more clear in Fig. \ref{fig:main}. Here, the integrated elliptic flow of charged hadrons is plotted at all three investigated values for $\epsilon_{\text{switch}}$ for the default choice of $\eta/s$ as well as the parameterizations in $T$ and $\epsilon$. We see that the lines for constant $\eta/s$ and $\eta/s(T)$ differ strongly for changing values of $\epsilon_{\text{switch}}$ - with reducing $\epsilon_{\text{switch}}$, the flow strongly increases. In contrast, for $\eta/s(\epsilon)$, the lines stay virtually on top of each other, which in turn means that the flow is, for this range of $\epsilon_{\text{switch}}$, independent of the switching energy density for $\eta/s(\epsilon)$.

As mentioned, the elliptic flow is strongly sensitive to the shear viscosity. When increasing $\epsilon_{\text{switch}}$, parts of the late stage evolution which were before performed in the hydrodynamic evolution, where the shear viscosity is defined by the input parameterization, are then performed in the hadronic transport, where the shear viscosity is not directly accessible.

This, in turn, means that the independence of the integrated flow from the value of $\epsilon_{\text{switch}}$ for $\eta/s(\epsilon)$ is a strong sign that $\eta/s(\epsilon)$ approximates the shear viscosity in the non-equilibrium hadronic transport stage. To phrase this important finding differently, the independence of the elliptic flow on the value of the switching transition energy density gives confidence that the viscous hydrodynamic and transport descriptions are equivalent over a broad region in the phase diagram.

\section{Conclusions and Outlook}\label{sec:Conclusion}
In this work, a new parameterization of the shear viscosity over entropy ratio as a function of energy density and net baryon number density, $\eta/s(\epsilon,\rho)$ based on known constraints is tested within the hybrid approach {SMASH-vHLLE-hybrid}. The parameterization is compared to a temperature dependent parameterization extracted from Bayesian inference as well as with the default setting of {SMASH-vHLLE-hybrid}, a constant value of $\eta/s$. Additionally, the impact of a $\rho$ dependence was investigated. For this comparison, the midrapdity yield and midrapidity mean transverse momentum of central collisions and the integrated elliptic flow of collisions at 20\% -- 30\% centrality at $\sqrt{s_{NN}}$= 7.7 GeV, 17.3 GeV and 39.0 GeV were studied.
The dependence on the net baryon number number does have no significant effect on the flow as its effect on the shear viscosity is limited to the early stages of the hydrodynamic evolution. Nevertheless, it introduces a significant difference in the yield and mean transverse momentum. In comparison with alternative parameterizations, $\eta/s(\epsilon)$ reproduces proton midrapidity yield and mean transverse momentum well, but underestimates the elliptic flow.
In a further study, it is shown that significant amounts of the elliptic flow originate from the hadronic transport in the initial condition and the rescattering phase. This is dependent on the collision energy, as this increases the lifetime of the fireball in the hydrodynamic evolution, as well as on the switching energy density. The flow is sensitive to the switching energy density when using constant or temperature dependent $\eta/s$, but stays at the same values for $\eta/s(\epsilon)$. This shows that the parameterization in the energy density approximates the shear viscosity in non-equilibrium hadronic transport within the region of $\epsilon$ = 0.1--0.5 GeV/fm$^3$.

For this study, bulk viscosity is neglected, but the effect of bulk viscosity on observables is expected to be significant. Since it is expected to peak around the phase transition, a parameterization of the bulk viscosity in the energy density is worthwhile to explore, too. Additionally, the introduction of non-constant viscosities affects the evolution of the fireball and can lead to a change in the effective shear viscosity. A higher bulk viscosity could, for example, slow down the expansion of the medium, which leads to higher energy densities and reduced shear viscosity.
Additionally, the range of collision energies has been restricted to up to 39.0 GeV in order to study regions where $\mu_B$ is high enough to see significant effects. The presented results show that, although the difference in the effective shear viscosity due to a non-vanishing $S_\rho$ are decreasing, the difference in midrapidity yield and mean transverse momentum are still significant due to the increased fireball lifetime. Therefore, it might be promising to include higher beam energies in the future. Additionally, increased statistics could give access to $v_3$, which has a considerably increases sensitivity to $\eta/s$.

\begin{acknowledgments}
This work was supported by the Deutsche Forschungsgemeinschaft (DFG, German Research
Foundation) – Project number 315477589 – TRR 211. N.G. acknowledges support by the Stiftung Polytechnische Gesellschaft Frankfurt am Main as well as the Studienstiftung des Deutschen Volkes.  Computational resources have been provided by the GreenCube at GSI. 
\end{acknowledgments}
\bibliography{2022_viscosity}
\end{document}